\documentclass[12pt, superscriptauthors]{article}
\usepackage{graphicx} 
\usepackage[letterpaper, margin=0.98in]{geometry}
\usepackage{hyperref}
\usepackage[sorting=none]{biblatex}
\usepackage[T1]{fontenc}
\usepackage{helvet}
\usepackage{titlesec}
\usepackage{authblk}
\usepackage{enumitem}
\usepackage{fancyhdr}

\makeatletter
\renewcommand\maketitle
  {
   {\centering\large\bfseries\@title}%
   \medskip\par\noindent
   {\centering\small\@author}%
   \hfill\par
  }

\titleformat{\section}[hang]
{\normalsize\bfseries}
{\thesection.}{0.5em}{}

\titleformat{\subsection}[hang]
{\normalsize\itshape}
{\thesubsection.}{0.5em}{}

\usepackage{titlesec}
\titlespacing*{\section}{0pt}{0.3\baselineskip}{0.2\baselineskip}
\titlespacing*{\subsection}{0pt}{0.3\baselineskip}{0.2\baselineskip}
\titlespacing*{\subsubsection}{0pt}{0.3\baselineskip}{0.2\baselineskip}
\titlespacing*{\paragraph}{0pt}{0.3\baselineskip}{0.2\baselineskip}

\title{Flexible Stellarator Physics Facility}

\author[1]{F.I. Parra\textsuperscript{a,}}
\author[2]{S.-G. Baek}
\author[1]{M. Churchill}
\author[3]{D.R. Demers}
\author[4]{B. Dudson}
\author[1]{N.M. Ferraro}
\author[5]{B. Geiger}
\author[1]{S. Gerhardt}
\author[1]{K.C. Hammond}
\author[1]{S. Hudson}
\author[5]{R. Jorge}
\author[1]{E. Kolemen}
\author[6]{D.M. Kriete}
\author[7]{S.T.A. Kumar}
\author[8]{M. Landreman}
\author[9]{C. Lowe}
\author[6]{D.A. Maurer}
\author[1]{F. Nespoli}
\author[1]{N. Pablant}
\author[5]{M.J. Pueschel}
\author[9]{A. Punjabi}
\author[1]{J.A. Schwartz} 
\author[7]{C.P.S. Swanson}
\author[5]{A.M. Wright}

\affil[1]{Princeton Plasma Physics Laboratory, Princeton, NJ}
\affil[2]{Plasma Science and Fusion Center, Massachusetts Institute of Technology, Cambridge, MA}
\affil[3]{Xantho Technologies, Madison, WI}
\affil[4]{Lawrence Livermore National Laboratory, Livermore, CA}
\affil[5]{University of Wisconsin - Madison, Madison, WI}
\affil[6]{Auburn University, Auburn, AL}
\affil[7]{Thea Energy, Princeton, NJ}
\affil[8]{University of Maryland, College Park, MD}
\affil[9]{Hampton University, Hampton, VA}

\date{}

\fancypagestyle{specialfooter}{%
\fancyhf{}
\fancyfoot[L]{
\textsuperscript{a}\small\selectfont ~Corresponding author: \href{mailto:fparradi@ppppl.gov}{fparradi@pppl.gov}}
\fancyfoot[C]{\thepage}
}

\begin{document}
\begin{center}
\centering
\maketitle
\end{center}

\thispagestyle{specialfooter}

\vspace{-15pt}

\section{Research objectives}
\label{sec:mission}

\emph{We propose to build a Flexible Stellarator Physics Facility to explore promising regions of the vast parameter space of disruption-free stellarator solutions for Fusion Pilot Plants (FPPs).}

The FESAC Long Range Plan recognized the quasi-symmetric stellarator as ``the leading US approach to developing disruption-free, low-recirculating-power fusion configurations'' \cite{power20}. To deliver the ambitious Decadal Vision for Commercial Fusion Energy, we must establish a persuasive stellarator program in parallel to the tokamak one: the stellarator will prove to be a better path to a reactor if theoretical predictions are confirmed and novel optimization techniques and strategies work as desired. Since the release of the Long Range Plan in 2020, stellarators have arguably made the most significant advances of all fusion concepts.  Groundbreaking results from W7-X demonstrated low neoclassical transport \cite{beidler21} and the successful operation of the island divertor \cite{jakubowski2021a}. Advances in theory and modeling now allow us to minimize turbulent transport \cite{robergclark23, jorge23, kim23}, to achieve equilibria with precise quasisymmetry \cite{landreman22}, to reduce neoclassical transport and fast ion loss to levels far below what has been previously achieved \cite{bader19, sanchez23}, and to minimize the effect of coil manufacturing errors \cite{wechsung22, Jorge_2023}.  If realized, these advances will lead to cost-effective stellarator designs with confinement comparable to tokamaks but without the fundamental challenges of disruptions and current drive.

At present, no planned or existing facility in the world can test these theoretical advances at the necessary scale (see section~\ref{sec:context}).  
For this reason, we propose a new, flexible mid-scale stellarator user facility for confinement and divertor studies that will validate these theoretical advances and deliver the physics basis needed by the fusion industry:
\begin{itemize}[leftmargin=1.5em, noitemsep, nolistsep]
    \vspace{0.25em}\item \textbf{Confinement.} In this area, the new facility's mission is
    \begin{itemize}[leftmargin=1em, noitemsep, nolistsep]
        \item Understand and control turbulent transport in a variety of 3D magnetic geometries; 
        \item Demonstrate the reduction of thermal and fast ion losses and of impurity accumulation to acceptable levels; and
        \item Enable the exploration of quasi-symmetric configurations at high ion temperature.
    \end{itemize}
    
    \vspace{0.25em}\item \textbf{Divertors.} The mission of the divertor in the proposed facility is
    \begin{itemize}[leftmargin=1em, noitemsep, nolistsep]
        \item Test the physics basis of the potentially revolutionary non-resonant divertor \cite{punjabi2020}; and
        \item Robustly handle power and particle exhaust and retain impurities across a range of core configurations. 
    \end{itemize}

    \vspace{0.25em}
    
\end{itemize}

\emph{These objectives are connected through the ultimate goal of demonstrating core-edge integration.} While this proposal is focused on these two highest priority objectives, the facility would also be able to address secondary objectives, such as finite-beta MHD stability limits. 

\section{Description}
\label{sec:description}

The design strategy for the proposed facility will focus on specific scientific gaps on the stellarator path to a disruption-free, steady state FPP: energetic particle confinement, turbulent ion transport and core/divertor optimization. The facility will also demonstrate the effectiveness of quasi-symmetry, the most promising strategy to minimize both neoclassical losses and energetic particle transport. Quasi-symmetry was first validated in the pioneering HSX experiment \cite{gerhardt05, Canik_2007}, but has not been demonstrated in larger, hotter plasmas relevant to FPPs, and it cannot be tested in existing facilities such as W7-X.  

Mid-scale machines cannot simultaneously match core (let alone edge) parameters ($\beta$, $\nu^*$, and $\rho^*$) for its FPP counterpart \cite{whyte2012}. We focus on reactor relevant $\nu^*$, and $\rho^*$, and acknowledges that this results in $\beta$ values that may be lower than a reactor may need. To deliver its confinement mission, the facility should target $\rho^* < 10^{-2}$ (crucial for the demonstration of energetic particle confinement), $\nu^* < 10^{-2}$ (so that stellarator neoclassical transport is in the $1/\nu$ or lower collisionality regimes, and Trapped Electron Modes, TEMs, can grow), temperatures above 1 keV and a minor radius of the order of half a meter to prevent excessive neutral penetration. By reducing the confinement mission to understanding and controlling turbulent transport by itself and not in combination with neoclassical and energetic particle losses, and the divertor objectives to testing the non-resonant divertor strike-line geometry only, both $\rho^*$ and $\nu^*$ can be increased, resulting in a smaller device that would need tailored wall materials or other strategies to reduce the effect of neutrals.  

We propose a two stage approach to the facility:


\begin{itemize}[leftmargin=1.5em, noitemsep, nolistsep]
    \vspace{0.25em}\item \textbf{Stage 1: Exploration -- Flexible magnetic configurations}: Explore magnetic configurations to optimize for minimum turbulent transport, validate fast-ion confinement, and study  plasma transport physics in non-resonant divertor configurations.

    \vspace{0.25em}\item \textbf{Stage 2: Exploitation -- High-power core-edge integration}: With candidate optimized configuration(s) from Stage 1, upgrade auxiliary power and device power handling capabilities to confirm desired core transport properties, compatible with divertor particle and power exhaust solutions.

    \vspace{0.25em}
\end{itemize}
This two-stage approach is similar in spirit to previous mid-scale stellarator white papers \cite{bader19}, though with more emphasis on a wider range of magnetic configurations.

\subsection{Stage 1: Exploration -- Flexible magnetic configurations}

The focus of Stage 1 is to provide the configuration flexibility that is needed to validate the recent advances in our understanding of transport in 3D fields, and to provide a testbed for scientific discovery in confinement physics. As such the focus of the design will be on magnetic and actuator flexibility to identify magnetic configurations that achieve key optimization targets of i) high neoclassical confinement, ii) good fast ion confinement and iii) turbulence optimization. 
We will assess a range of quasi-symmetric configurations to target, including the feasibility and cost of achieving both Quasi-Axisymmetry (QA) and Quasi-Helical-symmetry (QH) in a single device.

The flexibility described here has historically been achieved via a base configuration of modular coils with varying currents supplemented with smaller coils to trim the field (e.g. \cite{lee2022}). While this approach remains viable, we will also explore alternative means of achieving flexibility, with options including: (1) 2D continuous metal coils configurable with multiple voltage leads \cite{Kolemen_2024a, Kolemen_2024}, (2) dipole magnet setups on modifiable surfaces to be configured during major maintenance upgrades, and (3) multiple window pane coils individually controlled. The final choice will be made balancing technical maturity with potential benefits, with a preference for proven technologies that can deliver the mission.

The flexibility of this facility should extend to heating (anticipated to be mostly by neutral beams initially) and fuelling. Diagnostics will focus on the key profiles, the confinement of fast ions (e.g. neutral particle analyzers or loss probes), and ion scale turbulence diagnostics.



A primary mission of this facility is to assess the behavior of non-resonant divertors. Unlike other stellarator divertors (see section~\ref{sec:context}), non-resonant divertors are predicted to be resilient to changes in the plasma equilibrium \cite{bader2017}, a property currently being investigated in CTH. The next step is testing non-resonant divertors in a device with appreciable bootstrap currents and high-recycling divertor operation. Focus areas include stability of strike-point locations and variations in heat flux concentration as a function of equilibrium. 
Boronization or lithium wall conditioning may be employed to control density and edge impurities. We anticipate that for this stage of operation, a larger vacuum chamber and open divertor configuration will facilitate evaluation of the key physics. 

While this paper does not offer a definitive operating point, we have explored representative global parameters which could achieve the mission. We employ a simple 0D power balance analysis which uses the ISS-04 confinement time scaling, similar to the model in \cite{Alonso_2022}. As an example operating point, a device with $R=3$ m, $P_\mathrm{inj}=6$ MW (likely neutral beam heating of appropriate voltage), $A=5$, and $B =2.5$ T yields $n_{e,0} = 1.5\times10^{20}$ m$^{-3}$, $T_{i,0} = 3.5 \mathrm{keV}$,  $\beta=1.3\%$,  $\nu^*=2-5 \times 10^{-3}$ and $\rho^*=3-6 \times 10^{-3}$, and appears to be an appropriate point from which further optimization can be pursued. A magnetic field of 2.5 T facilitates flexibility -- it also enables the use of existing 140 GHz gyrotron technologies if ECRH is considered of interest, but this is not a strong constraint at present. The energy confinement time of $\sim 0.2$ seconds at these parameters implies a pulse length of at least $10\tau_E = 2.0$ seconds, setting minimum requirements for the wall armor and heating systems at this operating point. If the mission is de-scoped to be turbulence and divertor geometry studies only, a pathfinder machine with 1.7 T, $R = 2$ m and 6 MW can reach $n_{e,0} = 10^{20}$ m$^{-3}$, $T_{i,0} = 2.8$ keV, $\rho^* \sim 10^{-2}$ and $\nu^* \sim 1$. The conceptual design process would further refine the physics requirements and machine parameters.

\subsection{Stage 2: Exploitation -- High-power core-edge integration}

Stage 2 will entail upgrades to the heating and power handling capabilities of the facility, with the goal of exploiting the physics understanding from Stage 1 to study integrated core/edge solutions with simultaneous good core confinement and control of core impurity accumulation and scrape-off-layer (SOL) transport. The design of a divertor with high particle exhaust performance requires a known strike-line geometry, and therefore leads to a limitation on flexibility in magnetic configuration at high power. For this stage, we will select one or two optimized and validated magnetic configurations from Stage 1 for further development of the divertor concept.

Upgrades of the facility during Stage 2 will focus on divertor modifications, including closed divertor designs, new Plasma Facing Components (PFCs), baffling and targets, and active pumping (cryopumps or potentially a lithium system). Demonstration of good core confinement together with FPP-relevant wall material is an important next step in developing the stellarator path to fusion energy \cite{fellinger2023}. It is our preference to utilize inertially cooled high-$Z$ PFCs for Stage 2 studies in order to fully address the relevant impurity physics, though other materials will be considered, and the final choice will be informed by Stage 1. Multiple iterations of the divertor structures may be necessary to complete the Stage 2 mission, and modularity of design will be a key focus. If the non-resonant divertor proves to be as robust in practice as predicted, these iterations of the structure may be of modest scope.

The diagnostics for Stage 2 will need to advance to support the evolving mission. The divertor structures will be diagnosed for measurements of particle and heat fluxes, baffle pressures, and material erosion. The diagnosis of core impurity content will be enhanced. These measurements will allow a comprehensive exploration the physics of of core-edge integration, supporting both public and private sector research initiatives.

As an example operating point, the $R=3$ m device with an increased heating power of 14 MW attains ion temperatures of 3.9 keV and maintains the low values of $\nu^*$ and $\rho^*$ with  $\beta=2.5\%$, while increasing available heating power for challenging divertor systems.

\section{National and international context} 
\label{sec:context}

\subsection{The US program}


In section~\ref{sec:mission} we have listed some of the most impressive recent successes of the US stellarator theory program that justify the construction of this mid-scale facility. The cutting-edge experiments of the US stellarator program, such as the Helically Symmetric experiment (HSX) or the Compact Toroidal Hybrid (CTH), are at the forefront of the field, having validated quasi-symmetry \cite{gerhardt05, Canik_2007} and having confirmed the absence of disruptive MHD events in 3D fields \cite{archmiller14, pandya15}. These US-based machines contribute with substantial expertise and insights in terms of the design, construction and operation of stellarator experiments. The US program also has a strong presence in W7-X through the installation and operation of advanced diagnostics. The user facility that we propose is complementary to these programs.

\subsection {The international stellarator program}

A new mid-scale US stellarator has the potential to fulfill a unique and important role within the international stellarator program. Two major stellarator facilities are currently in operation: the Large Helical Device (LHD) in Japan, and W7-X in Germany. LHD has been in operation since 1999 \cite{iiyoshi1999a} and is scheduled to be permanently decommissioned in 2025. W7-X has been active since 2015 \cite{pedersen2017a} and anticipates many years of future operation. Both devices can run long pulse discharges (30+ minutes) at magnetic fields of 2.5--3 T. 

The proposed new US stellarator facility would immediately take on a global role as the most advanced stellarator and one of only two operating major stellarator facilities in the world after the shutdown of LHD. The cost of this new facility will be lower than the cost of W7-X because of advancements in coil design and because it can deliver its mission without long-pulse high-power operation, a technological challenge already addressed by W7-X. 

The objectives of the mid-scale US stellarator proposed in this paper focus on critical research areas that cannot be addressed by W7-X due to its particular design: 
\begin{itemize}[leftmargin=1.5em, noitemsep, nolistsep]
    \vspace{0.25em}\item \textbf{Confinement.} W7-X recently provided experimental confirmation of the success of its optimization strategy (different from quasi-symmetry) to reduce neoclassical losses \cite{beidler21}. However, its ability to confine energetic particles is inferior to that of more recent designs \cite{landreman22, bader19, sanchez23}, and its performance is limited by turbulent transport that clamps the ion temperature \cite{beurskens2021a}. While turbulence in W7-X can be suppressed through careful plasma profile shaping \cite{baldzuhn2020a, bozhenkov2020a, carralero2021a}, a steady-state, high-power, high-performance scenario has yet to be developed. Many advances have been made in stellarator optimization for turbulence reduction \cite{mynick2010a, xanthopoulos2014a, proll2016a, jorge23, kim23}; with a new mid-scale stellarator, we can test new plasma equilibria that reduce turbulent and energetic particle losses. 

    \vspace{0.25em}\item \textbf{Divertors.} The W7-X divertor has capably handled large heat loads and facilitated detached discharges \cite{jakubowski2021a}, but its particle exhaust capabilities to date have been weaker than comparable tokamaks \cite{wenzel2022a}. In addition, the W7-X divertor concept, known as an island-divertor, is generally only compatible with a limited class of stellarator configurations that have the property of low bootstrap current.  Due in part to its water cooling infrastructure required for long pulse operation, upgrades of the W7-X divertor take several years to complete, slowing the ability to qualify improved designs. The divertor of the proposed mid-scale device, by contrast, will be easier to replace, leading to more rapid design iterations tailored to different magnetic configurations.
\end{itemize}

\subsection{Private industry}

At present, there are at least eight private fusion companies advancing the stellarator concept \cite{typeoneenergy_web, theaenergy_web, stellarex_web, proximafusion_web, gaussfusion_web, helicalfusion_web, renaissancefusion_web, nttao_web} with a declared a total investment of over \$120M in funding altogether~\cite{FIA_2023}.  Of the three of these companies founded within the US (Type One Energy, Thea Energy, and Stellarex), the first two have been selected to participate in the DOE’s Fusion Milestone Program.

A public user-facility mid-scale stellarator would provide a national research program to leverage expertise across universities, national laboratories and private industry, flexibility to pursue a broad research program, safe experimental access, and an ideal environment to foster workforce development across a breadth of disciplines including at the Ph.D level.  The facility would be an attractive opportunity to leverage public investment with private support at all stages of design, construction and operation. That companies may propose experiments with similar research missions only enhances such opportunities for collaboration. We are supportive of public-private partnerships as long as they prioritize open science.

\section{Readiness of the facility}

The proposed facility is at the pre-conceptual (pre CD-0) level. The physics understanding needed to design the facility exist and is mature, as well as the tools needed to develop a detailed engineering design (see section~\ref{sec:mission}).  We have identified two extreme operating points: one targeting neoclassical, turbulent and energetic particle transport simultaneously, and non-resonant divertor performance; and the other focusing on turbulence and non-resonant divertor strike-line physics only. Both extremes can be achieved with copper or LTS magnets and with proven heating systems (gyrotrons and neutral beams) within the stipulated budget (\$180M -- \$720M), although a final decision on the size and magnetic field of the facility can only be made after a more thorough design effort. We categorize the readiness of this facility as "(b) significant scientific/engineering challenges to resolve before initiating construction."  Technical readiness is high with several optimized stellarator experiments having been successfully built (notably HSX, W7-AS, and W7-X) and several pre-conceptual, but detailed, design studies having been completed (e.g. \cite{bader2020a}). Aspects of technological demonstration could be made part of the mission scope in collaboration with private industry. 

\section*{Acknowledgements}
Prepared in part by PPPL under Contract DE-AC02-09CH11466 and LLNL under Contract DE-AC52-07NA27344. The United States Government retains a non-exclusive, paid-up, irrevocable, world-wide license to publish or reproduce the published form of this manuscript, or allow others to do so, for United States Government purposes.

\printbibliography

\end{document}